# Core to solar wind: a stepwise model for heating the solar corona

Claudio Vita-Finzi
Earth Sciences, Natural History Museum
Cromwell Road, London SW7 5BD, UK

*The model outlined here embodies three distinct, successive processes which both define and characterise the Sun's chromosphere, transition region and corona. Operating experience from fusion research shows how Spitzer resistivity may render ohmic heating in the chromosphere self-limiting and thus serve to define the lower margin of the transition region; its upper margin is at $\sim 6.10^3$ K, where radiative cooling of He/H plasma decelerates sharply. The third and last stage in the proposed scheme is expansion into the tenuous plasma of space, which leads to the acceleration of ions to high energies, long recorded by spacecraft instruments as $He^{++}$. There is thus dynamic continuity all the way from the solar interior - the energy source for spinning columns in the Rayleigh–Bénard setting of the convection zone - to the coronal exhalation of the solar wind, a finding which should benefit the analysis of space weather, witness the association between helium in the solar wind and the incidence of coronal mass ejections.*

The high temperature ($\geq 1\text{-}2.10^6$ K) of the Sun's outermost atmosphere or corona was identified in 1939 but has still to be explained. The mechanisms currently most in favour emphasise magnetic reconnection or waves of some kind and they treat the chromosphere and corona together (Amari et al 2015). This paper develops an alternative scheme (Vita-Finzi 2018) which links the Sun's interior with its atmosphere in three stages corresponding to (and indeed identifying) the photosphere-chromosphere, the transition region and the corona.

Bearing in mind that any analogy between processes on the Sun and in terrestrial laboratories -- particularly fusion (Morse 2018) -- is only approximate, there are instructive parallels between the first step in our model and the early stages of a conventional tokamak operation especially as laboratory experiments for these conditions are not available. There a toroidal current serves the dual purpose of confining the plasma and heating it. As the main contours of the solar body represent the interplay between gravitational contraction and thermal expansion, the solar environment performs confinement effectively though imperfectly, thus freeing the available magnetic energy from this task. In fact, as indicated by the solar wind, there is a net surplus of plasma to sustain the chromosphere.

**Step 1**

Plasma composition as well as induction heating shows qualified kinship between Sun and laboratory, although in a tokamak the favoured fuel – deuterium-tritium – is fully ionised at the temperatures required for fusion (c $10^8$ K). The H:He ratio may dominate discussion of the influence of elemental abundance on chromospheric heating, with a photospheric bulk composition of H 90.965% and He 8.89% (NASA 2018). Sodium, magnesium, calcium, and iron are also present, a fact that is exploited in particular in the assessment of fractionation between the photosphere and different varieties of solar wind (Peter & Marsch 1998). The impurities that have been detected during the ohmic heating phase of JET operation, such as reactor wall material (Ni, Cr, Fe), oxygen, carbon, molybdenum and chlorine, lead to radiation losses (Behringer et al. 1986) and presumably do so in the solar reactor.

The accepted view (NASA 2018) is that the temperature of the chromosphere rises from $66.10^2$ K at its contact with the photosphere to $\sim 3.10^4$ K over a distance of $\sim 25.10^5$ m. In our proposed tripartite scheme the weakly ionised Hα of the chromosphere is subject to ohmic (or Joule) heating. In accordance with the account by Spitzer (1958) the resistance and thus the efficacy of ohmic heating decrease in proportion to the electron temperature as $T_e^{-3/2}$., so that there is a point at which ohmic heating stalls. Owing to operational constraints (O'Brien & Robinson 1993) ohmic heating at startup in most tokamaks can attain at most ~ 1 keV , say $10^7$ K, as is the case with the JET tokamak (ESA 2013).

It has been suggested that the temperature of the chromosphere 'steadfastly refuses to rise above $10^4$ K until hydrogen becomes fully ionized' perhaps because 'ionization of hydrogen leads to a high specific heat' (Judge & Peter 1998, 190). The issue of specific heat had previously been raised in a study of the Jovian atmosphere for which an atmospheric composition of hydrogen and helium was postulated (Nelson 1971). A nondimensional plot of specific heat against temperature at $1\text{-}6.10^4$ K for particle densities from $10^{-10}$ to $10^{-6}$ g cm$^{-3}$ and for hydrogen unit volumes of 0.333 and 1.0 (equivalent to 50 % and of 100 % hydrogen by volume) yields two prominent peaks (Fig 1). The greater is at $2.5\text{-}4.10^4$ K, which may be manifested as a heightened but shortlived response to ohmic heating when the transiting gas attains a critical temperature. Thus specific heat imposes an upper limit on the chromospheric temperature well below the Spitzer limit. Indeed, the temperature in the Sun, after a temporary reversal, increases only to ~ $2.10^4$ K some $3.10^3$ km above the photosphere.

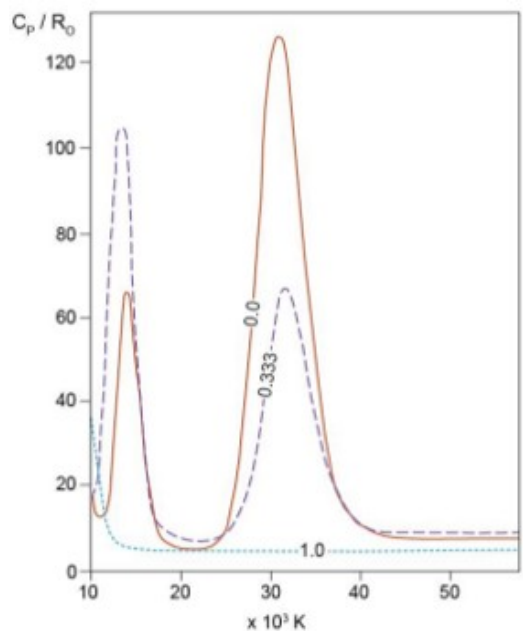

Fig 1 Plot of specific heat against temperature at $1\text{-}6.10^4$ K for particle densities from $10^{-10}$ to $10^{-6}$ g cm$^{-3}$ and for hydrogen unit volumes of 0.333 and 1.0 equivalent to 50 % and of 100 % hydrogen by volume (from Nelson 1971).

In our model of the Sun, induction is by way of electromagnetic energy derived from spinning convective pseudo-Taylor columns in the Rayleigh-Bénard setting of the convection zone -- pseudo in the sense that they may develop in a fluid subject to strong rotation and thermal forcing although without the basal obstacle of the original definition (Taylor 1921; Grooms et al. 2010; King & Aurnou 2012). Large-scale vortices are a possible outcome of rotating planar convection in an electrically conducting Boussinesq fluid (Guervilly et al. 2014). The associated dynamos generate magnetic fields that are concentrated in the shear layers surrounding the vortices, although for Rayleigh numbers just above a critical value the convection takes the form of elongated columns with a small horizontal cross-section and aligned with the rotation axis (Guervilly et al 2015). These are the structures that govern photospheric granulation (Vita-Finzi 2018).

The columnar model evidently differs from the classic notion of a primarily convective mechanism for granulation (e.g. November 1994). The summit of the columns is manifested as mesogranulation and supergranulation; the surface flow field is accordingly in close agreement with the magnetic field (Simon et al 1988). The columns are free to spin, even if closely packed, because they are insulated mechanically by sheaths (Sprague et al. 2006). Indeed, Spacelab-2 white-light images illustrate both clockwise and anticlockwise spin; they also show that photospheric vorticities can twist a magnetic flux tube by 360° in < 3 hr (Simon et al 1988), that is an average of >2°/min. Tangential (vortical) flows associated with the average supergranule outflow are indeed reported to reach about 10 m s$^{-1}$ (Langfellner et al. 2015).

The fluid uppermost photosphere in which they spin is partly ionised and therefore electrically conducting. The cylindrical support is irrelevant except insofar as it creates quasi-regular spacing of planar rotating discs at the photospheric surface. Large-scale-vortex dynamos, which call for magnetic Reynolds numbers ~100-550 (Guervilly et al. 2017; Bushby et al. 2018), are here proposed as the source of basal chromospheric heating. Analogy with the H/He atmospheric evolution of young terrestrial planets (Erkaev et al 2013) points to XUV radiation as a plausible supplementary heating source; XUV emission by the upper chromosphere and the TR was demonstrated by a slit spectrograph observation from Skylab (Doscheck et al 1975).

Magnetic energy flux at the photosphere has been evaluated at active regions, such as NOAA 11158 (Kazachenko et al. 2015), by modelling complemented by Hinode satellite observations. At one plage region the vertical Poynting flux had values of about $5\pm1 \times 10^7$ erg cm$^{-1}$s$^{-1}$ (Welsch 2015), close to the energy loss (~2 x10$^7$ erg cm$^{-2}$ s$^{-1}$) estimated for active-region fields in the chromosphere (Withbroe & Noyes 1977). The dominant heating mechanism, one of three discussed by Goodman (2000), is resistive dissipation of the proton (Pedersen) currents driven by the convection electric field that we have visualised as spinning columns.

Indeed, the modelling by Goodman (2004a) leads to the proposition consistent with the theme of this paper that the chromosphere of the Sun (away from flaring regions) is *created* by Pedersen current dissipation. Heating by Pedersen current dissipation is very inefficient when the plasma is fully ionized and strongly magnetised, somewhat above ~2170 km (Goodman 2004b), consistent with the value of 2500 km for the lower boundary of the Transition Region (NASA 2018) cited earlier.

Joule dissipation due to dynamo action is thought by Kan & Yamaguchi (1989) to account for heating in the photosphere-chromosphere and to amount in nonactive regions, on the basis of the classic quiet Sun model by Vernazza et al (1981), to ~2.4 x 10$^7$ erg cm$^{-2}$s$^{-1}$. The electromagnetic energy employed by induction may be approximated by the observed energy losses suffered, again during quiet Sun conditions, by the low chromosphere, which Withbroe & Noyes (1977) put at ~ $2 \times 10^6$ erg cm$^{-2}$ s$^{-1}$.

**Step II**

In a preliminary version of the tripartite scheme (Vita-Finzi 2018) the Joule-Thomson (J-T) effect was put forward as the pertinent heating system for the transition region (TR) though without the throttling that was included in the classic experiments by Thomson & Joule (1853). In the absence of experimental data for the temperatures at issue the term J-T is provisionally retained for the heating of a H/He plasma which is associated with a reduction in electron density $n_e$ from ~ $10^{19}$ to $10^{15}$ m$^{-3}$, that is to say when a strong negative density gradient in the quiet Sun coincides with a strong positive temperature gradient (Lemaire 1999).

In a widely reproduced diagram (Peter 2004; Fig 2) the onset of the TR corresponds to a plasma particle density $N$ (as distinct from 'plasma density' commonly used to signify electron density) of slightly more than $10^{16}$ m$^{-3}$.

Photoionisation of hydrogen reduces its cooling efficiency by some six orders of magnitude so that at high temperatures ($10^4$-$10^8$ K) neutral hydrogen cools at about $10^{-18}$ erg cm$^3$ s$^{-1}$ compared to $2.10^{-24}$ erg cm$^3$ s$^{-1}$ for ionised hydrogen, with a peak (to judge from the published data) at ~ $10^3$ K (Gnat & Ferland 2012). Photoionisation has a similar effect on helium (Oppenheimer & Schaye 2013), which when partially ionised cools very efficiently by blackbody radiation and direct coupling to the helium Lyman continuum. Once fully ionised by further heating, however, it no longer couples well to the continuum (the Lyman limit being 91.2 nm, 13.6 eV). This signals the end of radiative loss or, in other words, the onset of uninhibited heating, and temperatures of $10^6$ K are rapidly attained. In short the trigger is more in the nature of a safety catch which is released at the critical temperature.

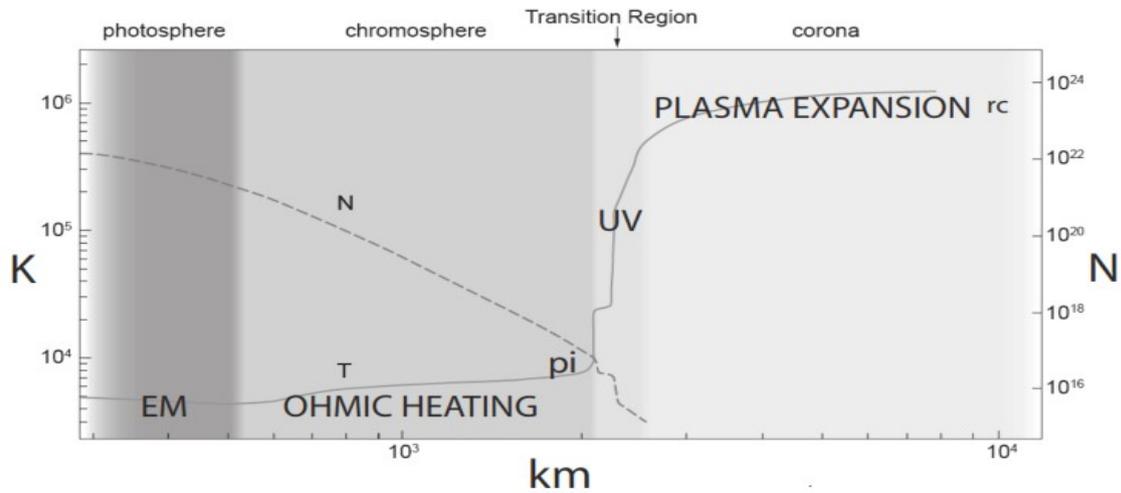

Fig 2 Proposed heating episodes and the intervening triggers set against major subdivisions of the solar atmosphere and plots of temperature (T) and plasma particle density (N). EM = electromagnetic energy, pi = photoionisation, rc = radiative cooling; T and N after Peter (2004).

A value of ~ $6.10^3$ K signals the region where cooling by radiation begins to nullify EUV heating as shown by radiative cooling functions for $^3$HeH$^+$ and $^4$HeH$^+$ (Coppola et al. 2011)(Fig 3). Here the rate of cooling attains between $10^{-10}$-$10^{-9}$ erg/s. Indeed the calculated radiative cooling function (in erg cm$^{-3}$ s$^{-1}$) at temperatures >$10^4$ K for plasmas at low densities with solar abundances in collisional ionisation equilibrium K drops rapidly from $10^5$ to $10^{7.5}$ K (Draine 2011).

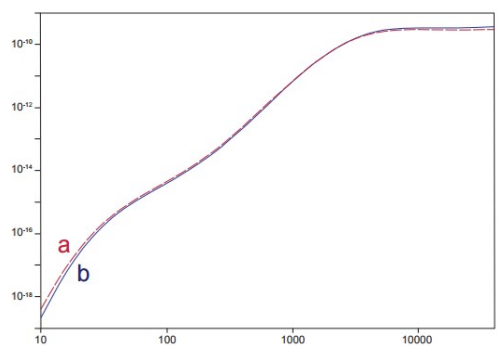

Fig 3 Radiative cooling function (erg/s) of HeH plasma (from Coppola et al. 2011). a = $^3$HeH$^+$, b = $^4$HeH$^+$

**Step III**

The upper limit of the TR may be defined about $5.10^6$ m above the photosphere, where the solar plasma has attained a value of $2.5 \times 10^5$ K, by a deceleration in the temperature increase then in progress. The onset of the TR corresponds to a plasma particle density $N$ of slightly more than $10^{16}$ m$^{-3}$ (Peter 2004; Fig 2). Photoionisation of hydrogen reduces its cooling efficiency by some six orders of magnitude so that at high temperatures ($10^4$-$10^8$ K) neutral hydrogen cools at about $10^{-18}$ erg cm$^3$ s$^{-1}$ compared to $2.10^{-24}$ erg cm$^3$ s$^{-1}$ for ionised hydrogen, with a peak (to judge from the published data) at $\sim 10^3$ K (Gnat & Ferland 2012). Photoionisation has a similar effect on helium (Oppenheimer & Schaye 2013). Thereafter heating, triggered by propinquity to the near-vacuum of space, continues equably in response to plasma expansion. Gurevich et al. (1966) and Gurevich & Pitaevsky (1975) were perhaps the first to show that the expansion of a plasma into a vacuum or a more tenuous plasma could result in the acceleration of ions to high energies (Samir et al. 1983), a process for which the self-similar solution indicates a logarithmic increase in velocity (Crow et al. 1975).

Plasma expansion has been investigated experimentally as well as theoretically (Chan 1986; Elkamash & Kourakis 2016) even though the circumstances that concern us here, viz. temperatures of $10^6$ K and coronal pressures of perhaps $1.3\ 10^{-11}$ Pa, present even more serious laboratory limitations than does the ohmic heating of plasmas in the chromosphere. But heating of He$^{++}$ ions in the solar wind has long been recorded by spacecraft (Ofman et al. 2015).

The bearing of this effect on space phenomena was made explicit by the interaction of an obstacle with a plasma. A relation between pressure fall and temperature in an astronomical context was assumed by Kothari (1938) when he showed that, for a relativistically degenerate gas (i.e. one nearing its ground state) undergoing Joule-Thomson expansion, the degree of heating per unit fall of pressure increased with the degree of degeneracy. Samir & Wrenn (1972) reported that ionospheric electron temperature measured by a Langmuir probe in the near wake of an artificial satellite (Explorer 31) was raised above that of the ambient electron gas by as much as 50 %. They referred to earlier work (Medved 1969) on the Gemini/Agena spacecraft in which wake temperature was 1700 K greater than the ambient temperature in one experiment and 764 K in another. The Moon's wake provided scope for related work; the increase in the electron temperature in the lunar wake found by the SWE plasma instrument on the WIND spacecraft amounted to a factor of four although ion temperatures were little changed (Ogilvie et al. 1996). Laboratory investigations based on immersion of a plate in a single-ion, collisionless, streaming plasma, saw 'early time expansion' result in ion acceleration into the wake (Wright et al. 1985).

**Conclusions**

Contrary to the accepted puzzling notion that the transition region and even the chromosphere are heated inwards from the corona (NASA 2018), the temperature rise is cumulatively radial. What is more, structuring of the solar atmosphere into three major zones is notthe source of our stepwise heting sequence but its outcome.

The coherence between solar wind variations and sunspot activity (Fig 4) is consistent with our proposed tripartite heating scheme: induction heating, which brings temperatures up to 20,000 K and triggers Joule-Thomson heating, which in turn results in temperatures of 250,000 K at the TR, and thereafter plasma expansion into the near vacuum of space, which is here proposed as the mechanism by which temperatures of 1-2 million K are raised in the corona before it grades into interstellar space.

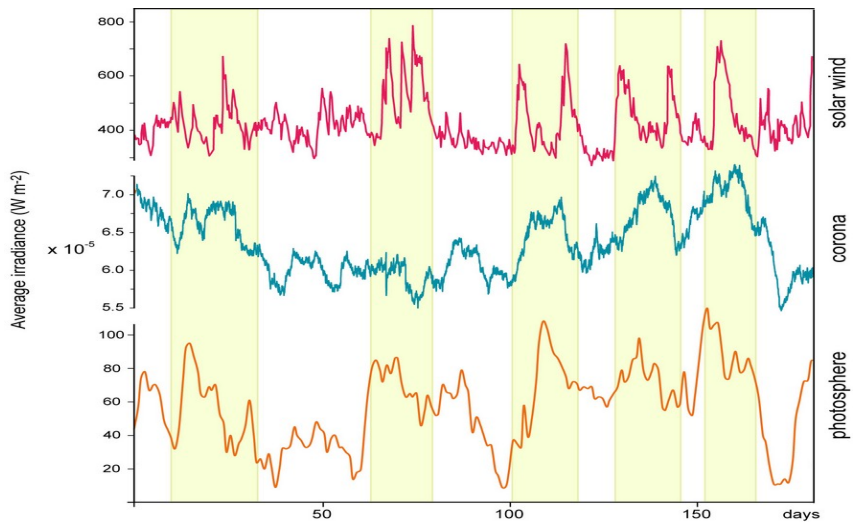

Fig 4  Irradiance variation for 1 Jan-1 July 2012 for photosphere, major subdivisions of the solar atmosphere, and solar wind . Plots and scale details (W m$^{-2}$) in Vita-Finzi 2018.

The long-term record of the Sun's activity, essential for robust interpretation of paleoclimates as well as for assessing the solar factor in weather, requires detailed information on the source of EUV fluctuations. Measurements by the EVE instrument on the Solar Dynamics Observatory satellite combined with neutrino data suggest that the UV flux is modulated primarily by rotation of the solar interior (provisionally named the Dicke Cycle:Vita-Finzi 2009) rather than the passage of active areas across the solar disc. Thus periodicities recorded by cosmogenic isotopes such as $^{10}$Be, which respond to oscillations in the strength of the solar wind, are better guides to the solar factor than observed sunspot records and have the advantage of spanning >$10^5$ yr rather than a mere $4.10^2$ yr. In short, the solar wind emerges as the one dependable indicator of solar activity. Sunspot data are compromised by their indirect relation to the Sun's irradiance: the rotation of active areas explains no more than 42% of its variation (Li et al 2012).

The proposed scheme could help to explain heating in other bodies (such as Titan) which display a radial increase in temperature and a decrease in plasma density as well as sustained gas outflow. It may also bear on the thermal evolution of other coronal stars.